\documentclass[fleqn,12pt,twoside]{article}
\usepackage{espcrc1}

\usepackage{graphicx}
\usepackage[figuresright]{rotating}

\newcommand{\AmS}{{\protect\the\textfont2
  A\kern-.1667em\lower.5ex\hbox{M}\kern-.125emS}}

\title{Nuclear structure of the exotic mass region along the {\it
rp} process path}
\author{Yang Sun \address[ND]{Department of Physics and
Joint Institute for Nuclear Astrophysics, \\
University of Notre Dame, Notre Dame, Indiana 46556, USA}, Michael
Wiescher \addressmark[ND], Ani Aprahamian \addressmark[ND], Jacob
Fisker \addressmark[ND]}

\begin{document}

% typeset front matter
\maketitle

\begin{abstract}
Isomeric states in the nuclei along the rapid proton capture
process path are studied by the projected shell model. Emphasis is
given to two waiting point nuclei $^{68}$Se and $^{72}$Kr that are
characterized by shape coexistence. Energy surface calculations
indicate that the ground state of these nuclei corresponds to an
oblate-deformed minimum, while the lowest state at the
prolate-deformed minimum can be considered as a shape isomer. The
impact of these shape isomer states on isotopic abundance in x-ray
bursts is studied in a multi-mass-zone x-ray burst model by
assuming an upper-lower limit approach.
\end{abstract}

\section{Introduction}

It has been suggested that in x-ray binaries, nuclei are
synthesized via the rapid proton capture process ({\it rp}
process) \cite{VanWorm,rpReport}, a sequence of proton captures
and $\beta$ decays responsible for the burning of hydrogen into
heavier elements. Recent reaction network calculations \cite{rp}
have shown that the {\it rp} process can extend up to the heavy
Sn-Te mass region. The {\it rp} process proceeds through an exotic
mass region with $N\approx Z$, where the nuclei exhibit unusual
structure properties. Since the detailed reaction rates depend on
the nuclear structure, information on the low-lying levels of
relevant nuclei is thus very important.

Depending on the shell filling, some nuclei along the {\it rp}
process path can have excited metastable states, or isomers
\cite{Nature}, by analogy with chemical isomers. It is difficult
for an isomeric state either to change its shape to match the
states to which it is $\gamma$-decaying, or to change its spin
orientation relative to an axis of symmetry. Therefore, isomer
half-lives can be very long. If such states exist in nuclei along
the {\it rp} process path, the astrophysical significance could be
that the proton-capture on long-lived isomers may increase the
reaction flow, thus reducing the timescale for the {\it rp}
process nucleosynthesis during the cooling phase.

Coexistence of two or more stable shapes in a nucleus at
comparable excitation energies has been known in nuclei with A
$\approx 70 - 80$. The nuclear shapes include, among others,
prolate and oblate deformations. In an even-even nucleus, the
lowest state with a prolate or an oblate shape has quantum numbers
$K^\pi$ = $0^+$. An excited 0$^+$ state may decay to the ground
0$^+$ state via an electric monopole (E0) transition. For lower
excitation energies, the E0 transition is usually slow, and thus
the excited 0$^+$ state becomes a ``shape isomer".

\section{Isomers in $^{68}$Se and $^{72}$Kr}

\begin{figure}
\includegraphics*[angle=0,width=21pc]{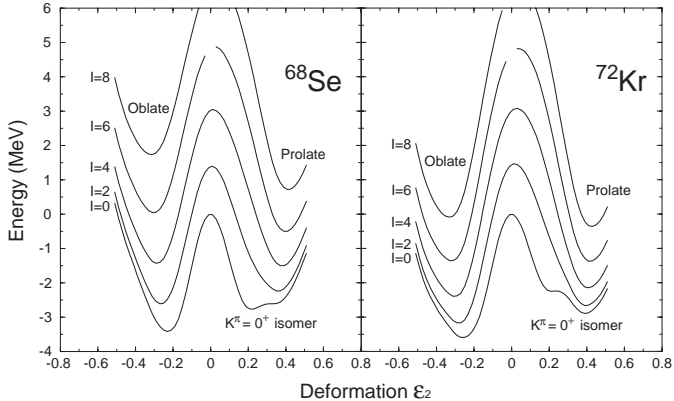}
\caption{Energy surfaces for various spin states as a function of
deformation variable $\varepsilon_2$.}
\end{figure}

Nuclear structure calculations are performed in the framework of
the projected shell model \cite{psm}. Fig. 1 shows calculated
total energies as a function of deformation variable
$\varepsilon_2$ for different spin states in $^{68}$Se and
$^{72}$Kr. The configuration space and the interaction strengths
in the Hamiltonian are the same as those employed in the previous
calculations for the same mass region \cite{Sun04}. Under these
calculation conditions, it is found that in both nuclei, the
ground state takes an oblate shape with $\varepsilon_2\approx
-0.25$. As spin increases, the oblate minimum moves gradually to
$\varepsilon_2\approx -0.3$. Another local minimum with a prolate
shape ($\varepsilon_2\approx 0.4$) is found to be 1.1 MeV
($^{68}$Se) and 0.7 MeV ($^{72}$Kr) high in excitation. Bouchez
{\it et al.} \cite{BouchezKr72} observed the 671 keV shape-isomer
in $^{72}$Kr with half-life $\tau = 38\pm 3$ ns. The one in
$^{68}$Se is our prediction, awaiting experimental confirmation.
Similar isomer states have also been calculated by Kaneko {\it et
al.} \cite{Kaneko04}.

In Fig. 2, we present the energy levels calculated by exact
diagonalization, and compare them with available experimental data
\cite{FischerPRL}. For $^{72}$Kr, with the newly confirmed $0^+$
isomer \cite{BouchezKr72} which should be the bandhead of the
prolate band, the rotational band at the prolate minimum is now
known. However, there have been no experimental data to compare
with the predicted oblate band. In contrast, an oblate band in
$^{68}$Se was observed and a prolate one was also established
\cite{FischerPRL}, except for the missing bandhead which we
predict as a shape isomer. For both nuclei, we predict low-lying
$K$-isomers, indicated by bold lines. In particular, the spin-16
states are so low in excitation (much lower than the spin-16 state
in the ground band) that one may consider them as an energy trap
\cite{Nature}.

\begin{figure}
\includegraphics*[angle=0,width=29pc]{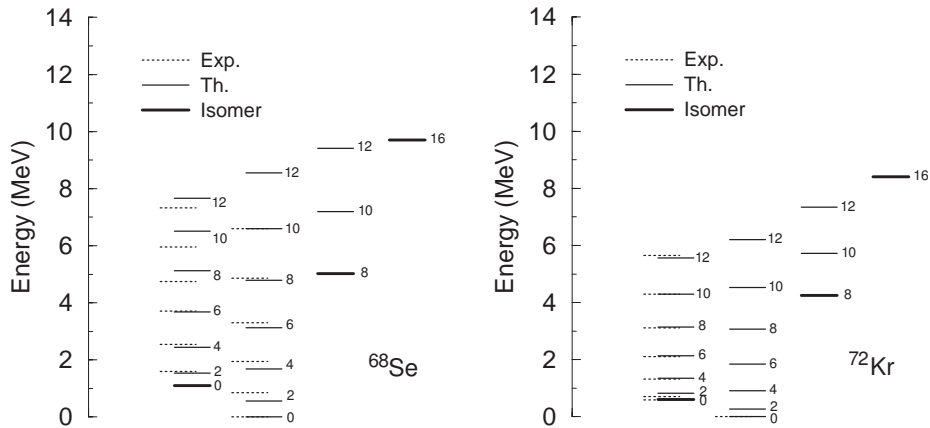}
\caption{Energy levels for $^{68}$Se and $^{72}$Kr.}
\end{figure}

\section{Possible impact on isotopic abundance in x-ray bursts}

The recent observation of a low energy 0$^+$ shape isomer in
$^{72}$Kr \cite{BouchezKr72} has opened new possibilities for the
rp-process reaction path. A similar shape isomer has been
predicted for $^{68}$Se in this paper. Since the ground states of
$^{73}$Rb and $^{69}$Br are bound with respect to these isomers,
proton capture on these isomers may lead to additional strong
feeding of the $^{73}$Rb($p,\gamma$)$^{74}$Sr and
$^{69}$Br($p,\gamma$)$^{70}$Kr reactions. However, whether these
branches have any significance depends on the associated nuclear
structure parameters, such as

%\begin{itemize}
%\item{
\noindent (1) how strong is the feeding of the isomer states?
%} \item{

\noindent (2) what is the lifetime of the isomer with respect to
$\gamma$-decay and also to $\beta$-decay?
%} \item{

\noindent (3) what are the lifetimes of the proton unbound
$^{69}$Br and $^{73}$Rb isotopes in comparison to the proton
capture on these states?
%}
%\end{itemize}

Two processes can be envisaged to populate the isomeric states in
appreciable abundance: either through thermal excitation of the
ground state at high temperatures, or through proton capture
induced $\gamma$-feeding. Thermal excitation is very efficient for
feeding levels with low excitation energy since the population
probability scales with $e^{-E_{is}/kT}$. Contributions of low
energy states (E$_x\le$ Q) are negligible since proton capture on
those states is balanced by inverse proton decay \cite{rpReport}.
This is not the case for proton capture on the isomeric states.
The peak temperature in the x-ray burst model employed in this
work is around 1.5 GK, and the isomer states in $^{68}$Se at 1.1
MeV and in $^{72}$Kr at 0.67 MeV are therefore only very weakly
populated with the probability $\le 0.02\%$ and $\le 0.5\%$,
respectively. Feeding through $^{67}$As($p,\gamma$)$^{68}$Se$^*$
(Q $\approx$ 3.19 MeV) and $^{71}$Br($p,\gamma$)$^{72}$Kr$^*$ (Q
$\approx$ 4.1 MeV) is a more likely population mechanism.
Nevertheless, a quantitative prediction of the feeding probability
requires a more detailed study of the $\gamma$-decay pattern of
low-spin (I $\le$ 3) states above the proton threshold in
$^{68}$Se and $^{72}$Kr, respectively.

The lifetime of the isomeric states must be sufficiently long to
allow proton capture to take place. No information is presently
available about the lifetime of the $^{68}$Se$^*$ isomer while the
55 ns lifetime of the isomer in $^{72}$Kr \cite{BouchezKr72} is
rather short. Based on Hauser-Feshbach estimates \cite{rpReport}
the lifetime against proton capture is in the range of $\approx$
100 ns to 10 $\mu$s, depending on the density in the environment.
Considering the uncertainties in the present estimates a fair
fraction may be leaking out of the $^{68}$Se, $^{72}$Kr
equilibrium abundances towards higher masses.

However, the process also depends on the actual proton-decay
lifetimes of $^{69}$Br and $^{73}$Rb. Based on model dependent
fragmentation cross-section predictions for these isotopes,
lifetimes have been estimated to be less than 24 ns and 30 ns,
respectively \cite{pfaff}. Again, within the present systematic
uncertainties this is in the possible lifetime range of proton
capture processes in high density environments.

\begin{figure}
\includegraphics*[angle=0,width=32pc]{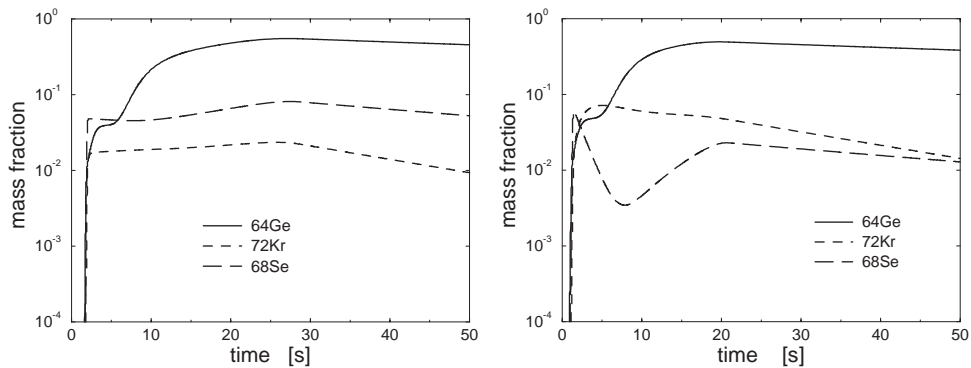}
\caption{Mass fractions in the x-ray burst model with two extreme
cases.}
\end{figure}

While it is likely that equilibrium is ensued between all these
configurations within the presently given experimental limits a
considerable flow towards higher masses through the isomer branch
cannot be excluded. Figure 3 shows the comparison between the two
extreme possibilities for the reaction sequence calculated in the
framework of a multi-mass-zone x-ray burst model \cite{fisker}.
The left-hand figure shows the mass fractions of $^{64}$Ge,
$^{68}$Se, and $^{72}$Kr as a function of time neglecting any
possible isomer contribution to the flow. The right-hand figure
shows the results from the same model assuming full reaction flow
through the isomeric states in $^{68}$Se and $^{72}$Kr rather than
through the respective ground states. The main differences in
$^{68}$Se and $^{72}$Kr mass fractions are due to rapid initial
depletion in the early cooling phase of the burst. This initial
decline is compensated subsequently by decay feeding from the long
lived $^{64}$Ge abundance. The results of our model calculations
shown in Fig. 3 are based on upper and lower limit assumptions
about the role of the shape isomer states. The possible impact on
the general nucleosynthesis of $^{68}$Se and $^{72}$Kr turns out
to be relatively modest. These assumptions are grossly simplified.
Improved calculations would require better nuclear structure data
to identify more stringent limits on the associated reaction and
decay rate predictions.

\end{document}